# Nano scale phase separation in Au-Ge system on ultra clean Si(100) surfaces

A. Rath[1], J. K. Dash[1], R. R. Juluri[1], Marco Schowalter[2], Knut Mueller[2], A. Rosenauer[2] and P. V. Satyam[1,*]

[1] Institute of Physics, Sachivalaya Marg, Bhubaneswar - 751005, India

[2] Institute of Solid State Physics, University of Bremen, D-28359 Bremen, Germany

**Abstract**

We report on the formation of lobe-lobe (bi-lobed) Au-Ge nanostructures under ultra high vacuum (UHV) conditions ($\approx 3\times 10^{-10}$ mbar) on clean Si(100) surfaces. For this study, $\approx$2.0 nm thick Au samples were grown on the substrate surface by molecular beam epitaxy (MBE). Thermal annealing was carried out inside the UHV chamber at temperature $\approx 500^\circ$C and following this, nearly square shaped $Au_xSi_{1-x}$ nano structures of average length $\approx$ 48 nm were formed. A $\approx$2 nm Ge film was further deposited on the above surface while the substrate was kept at a temperature of $\approx 500^\circ$C. Well ordered Au-Ge nanostructures where Au and Ge residing side by side (lobe-lobe structures) were formed. In our systematic studies, we show that, gold-silicide nanoalloy formation at the substrate (Si) surface is necessary for forming phase separated Au-Ge bilobed nanostructures. Electron microscopy (TEM, STEM–EDS, SEM) studies were carried out to determine the structure of Au – Ge nano systems. Rutherford backscattering Spectrometry measurements show gold inter-diffusion into substrate while it is absent for Ge.
.

# 1. Introduction

The importance of the study of surface and interface phenomena has been steadily increasing and now constitutes one of the most significant aspects of the semiconductor industry. Knowledge of surface properties of Si, Ge and Au nanostructures are crucial for rapidly developing area of nanotechnology. Metal nanoparticles on semiconductor substrate, facilitates to exploit the nano-particle properties and to fabricate novel and larger integrated devices. Metal nano-particles exhibit unique electronic, magnetic, photonic and catalytic properties which can be employed for the preparation of new materials for energy storage, photonics, communications, and sensing applications [1, 2]. For example, metal over layers have been exploited to tune the characteristics of epitaxial islands/ quantum dots, including their size, density and shape [3, 4]. Additionally, metallic films such as Au on Si have been instrumental in catalyzing the growth of semiconductor nano wires [5] and nano tubes [6, 7]. Starting in the early seventies, the Au–Si junctions in bulk or thin film samples have been extensively investigated using various surface techniques to understand their crystallographic, chemical, and electronic properties [8–13].

Epitaxially grown self organized nanostructures on silicon have been studied in great detail [14-17]. The size, the shape, and the material composition of nanostructures for the future prospective applications of nanostructures in optoelectronic and high density data storage devices have been fabricated using MBE facilities. Also, both the homo-epitaxial and the hetero-epitaxial growth of different metals on (100) oriented substrate has been studied by several groups [18, 19]. Recently, we reported the formation of aligned gold-silicide-nano rectangular structures, when a ~2nm Au film deposited on Si(100) in absence of native oxide at the interface using MBE method and was thermally annealed inside the TEM chamber (*in-situ* TEM)) [20].

Successful implementation of germanium technology will however require an understanding of the solid-state interaction in metal-germanium systems [21-23]. Ge is of particular interest because it has high mobility for both electrons and holes, making a high-speed complementary field-effect transistor a possibility [24]. Because of the low vapour pressure, Au-Ge alloys are used as catalyst for the growth of Ge nano wires inside the UHV chamber [25].

The effect of size on phase stability and phase transformations is of both fundamental and applied interest. For example, during the nucleation and growth of self-assembled nanowires from nanoscale metal catalysts, the phase of the catalyst determines properties such as the growth rate and the structure of the nanowire. Any size-dependent or growth-rate dependent changes in the catalyst may thus have strong effects on the structures that formed [26 – 32]. In last five years, interesting works have been carried out in understanding growth of silicon and germanium nanowire in presence of Au as catalyst and deviation in phase diagram of Au-Si and Au-Ge systems by using in-situ TEM methods [26 – 32].

Systematic results reported by Sutter et al. clearly point out the deviation of Ge solubility from bulk for nano-scale systems (typically less than 50 nm size) [27]. The systems dealt by Sutter et al., are basically for VLS type growth (free standing out-of-plane NW growth or encapsulated NW growth) and the results showed that the phase diagram would deviate from the bulk [27]. Similar results were shown by Kim et al. for Au – Si systems [30]. Cheuh *et al* reported on the post growth engineering of the nanowire (NW) structures and composition through the alloying and phase segregation that is induced by thermal annealing in Au-Ge system [33]. In these studies, the authors reported the dependence of the Au concentration of the initial nanowires on the formation of nearly periodic nanodisc patterns to core/shell and fully alloyed nanowires. Electrical measurements showed that alloy structures are having metallic characteristics [33].

In this article, we present the experimental observation of growth of aligned bilobed Au-Ge nano-structures during *in-situ* thermal treatment inside UHV chamber for gold deposited on Si(100) surfaces without oxide layer at the interface followed by further deposition of Ge on it. E*x-situ* electron microscopy measurements (both SEM and TEM) confirm the presence of Au and Ge in these bi-lobed structures. The role of gold silicide in formation of the bi-lobed structures has been clearly demonstrated in this work. Also, we have studied the effect of substrate temperature and the sequence in the material deposition on formation of bi-lobed structures. We also show the formation of Ge nanostructures of rectangular shape on the surface where no bi-lobed structures are present. This work will lead to the study of multi component materials, i.e., assemblies of mixtures of nanoparticles at single Silicon substrate.

**2. Experimental**

For this present study, we have prepared seven types of samples. For the first case (sample A), a thin Au film of thickness of about ≈2.0 nm on n-type Si(100) of resistivity 10–20 Ω cm, by MBE method under ultra high vacuum conditions [34]. Si (100) substrates were loaded into the MBE chamber and degassed at ≈ 600°C for about 12 hours inside the chamber and followed by flashing for about 2 minutes by direct heating at a temperature of ≈1200°C. In this process, native oxide was removed and a clean Si (100) surface was obtained. *In-situ* reflection high energy electron diffraction (RHEED) measurements confirmed that the Si(100) surface is 2×1 reconstructed (data not shown). On such ultra clean surfaces, ≈2.0 nm thick gold films were grown epitaxially by evaporating Au from a Knudsen cell. Deposition rate was kept constant at ≈ 0.14 nm min$^{-1}$. Table 1 gives an overall view about the specimen used in this work. Thermal annealing of the as-deposited sample (sample A) was carried out inside the UHV chamber at temperature of 500°C with a ramp rate of 7°C min$^{-1}$(sample B). For the third case (sample C), we deposited a ≈ 2.0 nm gold on Si(100) sample in the MBE chamber as explained above (like sample A) and about 2.0 nm Ge was further deposited at a typical deposition rate of 0.6 ML/min (where one ML corresponds to 6.78 × 10$^{14}$ atoms cm$^{-2}$ for a Si(100) surface) at a substrate temperature of 500°C. For fourth case (sample D), we have deposited 2.0 nm Ge on the Au patterned (after cooling the sample B to RT) surface while the substrate is *at room temperature* (RT). For fifth case, 2.0 nm Au was first deposited followed by 2.0 nm Ge, while the substrate was kept at RT. After the deposition of Au and Ge at RT, the specimen was annealed for 30 minutes at 500º C (sample E). For sixth case (sample F1, F2, F3), a 5.0 nm thick Ge film was deposited (instead of 2.0 nm) using same procedure as above on 2.0 nm Au deposited surface( like sample A) with various substrate temperatures: 400°C (F1), 500°C (F2) and 600°C(F3). During the growth, chamber vacuum was ≈ 6.2 ×10$^{-10}$ mbar. The post-growth characterization of the above samples were done with the field emission gun based scanning electron microscopy (FEGSEM) measurements with 20 kV electrons using a Neon 40 cross-beam system (M/S Carl Zeiss GmbH). Both cross-sectional and planar TEM specimens were prepared from the above samples in which electron transparency was achieved through low energy Ar$^+$ ion milling. For seventh case (sample G), As-deposited MBE grown 2.0 nm Au /Si(100) (like sample A) was *in-situ* annealed (up to 400°C) inside the TEM using single tilt heating holder (GATAN 628UHR, that has a tantalum furnace and can heat the specimen up to 1000°C) with the ramp rate of 7°C min$^{-1}$ . The temperature is measured by a Pt/Pt-Rh thermocouple and is accurate within a couple of degrees. The holder has a water cooling system to avoid over heating of the sample surroundings and the specimen chamber, while keeping only the sample at a specified temperature. The summary of all the above samples are presented in Table 1. TEM measurements were performed with 200 keV electrons (2010, JEOL HRTEM) and scanning TEM (STEM), high resolution TEM measurements were done using 300 keV electrons in the C$_s$-corrected FEI Titan 80/300 system. The Rutherford backscattering (RBS) measurements were carried out using 2.0 MeV He$^+$ ions.

## 3. Results and discussions

Figure 1(a) depicts a bright field TEM micrograph of as deposited 2.0 nm Au Film on Si(100) without native oxide at the interface (sample A). Irregular and isolated gold nanostructures of typical size/length of 29.3 ± 1.4 nm were formed. It should be noted that the error bars shown here is the error in determining the peak value *not* the distribution variation. The selected area diffraction pattern (SAD) taken on a group of nanostructures confirms the polycrystalline nature of the gold films (figure 1(b)). A high resolution (HR) TEM taken on a single isolated nanostructure shows the lattice spacing corresponding to elemental gold (inset figure 1(a)). The as-deposited sample was annealed *in-situ* inside the UHV chamber at 500°C for 30 minutes (sample B). Following this annealing in UHV, ex-situ TEM measurements (at room temperature) showed rectangles/squares nanostructures aligned along set of planes {110} direction with respect to the silicon substrate (Figure 1 (c)). This shape transformation is attributed to the strain relief mechanism [20]. Inset of figure 1 (c) depicts the bright field HRTEM image of one of the nano island. Figure 1(d) and 1(e) shows the corresponding selected area diffraction (SAD) pattern taken on group of rectangles/squares and cross-sectional HRTEM image of single rectangles, respectively. SAD shows the single crystalline silicon background and some weak polycrystalline rings of gold along with the dotted ring which corresponds to a d-spacing of about 0.226 nm which matches with the several phases such as $Au_5Si_2$, $Au_5Si$, $Au_7Si$ [35]. Figure 1(f) shows the SAD pattern taken at RT from the same sample but heated upto 400°C (sample G). There was no signal of silicide formation at 400°C. It should be noted that this annealing was carried out using *in-situ* TEM heating stage. More discussion on the effect substrate temperature which plays an important role in formation of gold-silicide nano alloy structures and helps to form Au-Ge bi-lobed structures will be discussed later sections in this paper.

The average ratio of length along <1 1 0> to length along < 1 -1 0> direction, known as "aspect ratio" was determined using several frames like figure 1(c). The length distribution along <1 1 0> shows the Gaussian distribution and the average length is 48.1± 1.1 nm (figure 2(a)). The average aspect ratio has been estimated by fitting the histograms with Gaussian distribution and the average value is 0.998 ± 0.001 indicating that the islands are predominately compact in shape (figure 2(b)). In our earlier work [20], we had shown that above 510°C, the average length (25.1 ± 1.6) and aspect ratio (1.12 ± 0.01) values were saturated during the *in-situ* thermal annealing inside the TEM (high vacuum chamber). The percentage of particles having rectangular/square shape was 80%. But here (in UHV annealing), one can clearly see that the almost all the particles are become squares/rectangles having decreased average aspect ratio and increased average length. When a ≈ 2 nm Ge was further deposited (sample C) using MBE system on the above rectangular/square patterned surface at substrate temperature 500°C inside the UHV chamber, bilobed nanostructures of Au and Ge were formed. It should be noted that AuSi nanalloys are used as seeding positions for Au-Ge lobe-lobe structures. This means that, Au deposition and annealing and Ge deposition after Au deposition are done in sequentially without any break in the vacuum. In figure 3(a), SEM image shows the bi-lobed nanostructures of Au and Ge, where the bright contrast is the gold and other one is for the Ge. As Au is higher Z material, more secondary electrons are emitted from the Au region than from Ge region and this causes the contrast difference between gold and germanium in SEM images. The average length and aspect ratio of the bi-lobed (including both Au and Ge) structures is 116 ± 1.8 nm and 1.12 ± 0.03 respectively. The length distribution is fitted with Gaussian distribution and the aspect ratio is fitted with log normal distribution to estimate the average value (figure 2(c)-(d)). In addition to the bi-lobed structures, one can see the formation of rectangular shaped Ge nanostructures having typical size ~ 3.0 nm (figure 3(a)). In figure 3(c), SEM images taken at 54° tilt have been shown. This shows a three dimensional nature of the Au-Ge nanostructures. Figure 3(b) and 3(d) depicts high resolution XTEM of the epitaxially grown Ge nano structures and Au-Ge bilobed structures on the silicon substrate respectively. It shows the diffusion of gold into the silicon substrate, where as there was not appreciable diffusion of Ge. This is further confirmed with Rutherford backscattering spectrometry (RBS) method. The RBS results would be discussed in the following sub-section.

We now discuss on the detail structure of bi-lobed structures and the interface between the Au lobe and Ge lobe. The bright field (BF) TEM image in figure 4(a) depicted one of the bilobed structures and the corresponding HRTEM image is shown in figure 4(b). At the Au–Ge junction (dotted circle) using high resolution images, the d-spacing has been found to be of about 0.332 nm and 0.279 nm. These lattice spacing do not match with the Au, Ge or Si but matches quite well with the (224) and (324) plane of tetragonal phase of $Au_{0.6}Ge_{0.4}$ [36]. According to bulk Au-Ge phase diagram, the above composition is formed at $520°C$ which is comparable with our growth temperature. Also, the lattice spacing value of 0.279 is close to (200) planes of $Si_{0.25}Ge_{0.75}$. We need to carry our more studies to come to a conclusion on the interfacial structure (i.e. whether it is AuGe or SiGe). It should be noted that deviations in the nano-scale phase-diagram to bulk phase diagram are reported by Sutter et al. [27].

Figure 4 (c), depicts a STEM– HAADF image showing the large area of the specimen with bi-lobed structures. This shows the expected Z-contrast, so that Au appears brighter than the Ge over layer and Si substrate. To analyze the local distribution of Au in these structures, STEM–x-ray energy dispersive spectrometry (XEDS) elemental mapping was carried out. In Figure 4(d), STEM EDS elemental mapping shows the presence of Au and Ge in a single bi-lobed structure. Though it is clear about the phase separation of gold (from 4(d) – Au M signal), but it is not very clear about the possible presence of SiGe. We need to carryout high resolution cross-section STEM measurements to obtain this information.

Recently, Gamalski et al have shown the in-situ study of formation of liquid Au-Ge alloy during digermane exposure at lower temperature (below 300 °C) [37]. In this work in-situ real time study has not been done to observe the growth of above nanostructure with temperature. To do the ex-situ temperature dependent study, we deposited 5.0 nm Ge on Au patterned surface at 400°C, 500°C and 600°C. Formation of bi-lobed structures was not found in 400°C case (figure 5 (a): sample F1). Whereas for 500°C and 600°C, more prominent bi-lobed structures were formed (figure 5(b), 5(c)). For 500°C (sample F2), the typical size of the structure was ~131 ± 1.7 nm. That means the average size of the bi-lobed structure increases with increasing the amount of Ge. In 600°C (sample F3) case, the mean size was 158 ± 2.3 nm which indicates that for same Ge thickness, average size increases with the temperature of the substrate. The Rutherford backscattering spectrometry measurements shown in Figure 5 (d), confirms the inter-diffusion of gold into Silicon and stability of Ge as a function of temperature. The inter-diffusion of gold into silicon is very well studied,

In the following, we discuss on the possible surface and interface energetic. We calculated the $\theta_1$ and $\theta_2$ values from the image. From that we calculated the value of $\gamma_1$ and $\gamma_2$ using the relation shown in the text ((Equation 1 and 2)). The notations are per the reference [38].

$$\gamma_1 \cos\theta_1 + \gamma_2 \cos\theta_2 = 1 \quad \text{---------- (1)}$$

$$\gamma_1 \sin\theta_1 = \gamma_2 \sin\theta_2 \quad \text{---------- (2)}$$

Where

$\theta_1$ is the angle between the Au lobe and the normal drawn at the Au, Ge and Si triple point

$\theta_2$ is the the angle between the Ge lobe and the normal drawn at the Au, Ge and Si triple point

$$\gamma_1 = \frac{\gamma_{Au/Si}}{\gamma_{Au/Ge}} \text{ and } \gamma_2 = \frac{\gamma_{Ge/Si}}{\gamma_{Au/Ge}}$$

(Where $\gamma_{Au/Si}$, $\gamma_{Ge/Si}$ and $\gamma_{Au/Ge}$ are the interfacial free energies for the Au-Si, Ge-Si and Au-Ge interfaces).

To check whether the bi-lobed structures are equilibrium structures or not, we calculated the $\gamma_1$ and $\gamma_2$ values for the bi-lobed structures and applied the formulas used by the Yuan et al [38] in their calculation. The value of the $\gamma_1$ (= 1.05) and $\gamma_2$ (= 1.07) were well matched with the studied done by Yuan et al [38]. So these Au-Ge bilobed structures are most likely the equilibrium structures. We have found that by changing the Ge thickness and working temperature, the size of these bilobed structures can be tuned. From the all above studies, it was found that the heated Au-Si alloy structures may play a major role in formation of such bilobed structures. During the deposition of Ge, gold silicide particles act as nucleation center for the growth of Ge. The minimum surface and interfacial energy for the system is achieved when the Ge is nested within Au and this would be a critical factor during the nucleation of the Ge crystals, where surface energy considerations determine the size and shape of the critical nucleus. As the deposition continues, the smaller liquid droplets provide a rapid diffusion path for the incoming Ge atoms, which find it energetically favorable to join already-nucleated crystals rather than to nucleate new ones. For larger thickness of Ge (5.0 nm), more Ge is nested within Au due to the enhancement of material availability. It resulted in increase in size of the Au-Ge bilobed structures. One can clearly see that at constant temperature of 500°C, the size of the Au-Ge bilobed structures was typically~ 116nm for 2nm Ge case whereas for 5nm Ge case it was ~131 nm.

To study the effect of substrate temperature, we deposited 2.0 nm Ge on the gold patterned substrate at room temperature (sample D). In figure 6(a), SEM image shows the formation of rectangular gold structures and Ge over layers in between the gold structures. Formation of bi-lobed structures was not found in this case. Also we deposited 2 nm gold and then 2 nm Ge on Si(100) substrate and heated it upto 500°C (sample E). During annealing, one would expect the formation of the regular structure following the substrate symmetry of Si(100) as seen in our previous work [20]. But here, both Au and Ge were present on the surface during annealing and as the thermal diffusion coefficient of both the materials is different [39], we observed the formation of irregular Au-Ge structures (figure 6(b)). Also, it has been reported that the $Au_xGe_{1-x}$ (0.27<x<0.6) shows super conductivity with a transition temperature up to ~ 2K [40]. There is a possibility to use the combination of Au, $Au_xSi_{1-x}$, Ge, and $Au_xGe_{1-x}$, structures. But, more studies are necessary to have a control on specific combination of the above structures so that technological applications become viable.

## 4. Conclusions

We have reported the formation of Au-Ge bilobed nanostructures by using molecular beam epitaxy. The effect of substrate temperature, material deposition sequence and also the role of gold silicide on the formation bi-lobed structures were studied in detail.

**Acknowledgements**

P. V. Satyam would like to thank the Department of Atomic Energy, Government of India, for 11th Plan Project No 11-R&D-IOP-5.09-0100 and the University of Bremen, Germany, for his sabbatical visiting program.

**Figure captions:**

**Figure 1:** (a) As-deposited (sample A) showing gold nanostructures with corresponding high resolution image of one of the islands which is showing the d-spacing of Au(111) in the inset (b) SAD pattern showing reflection of both Si and Au, (c) bright field TEM image taken for the sample that was annealed at 500°C in UHV chamber (sample B) with corresponding high resolution image of one of the single islands (inset), (d) SAD pattern showing the formation gold silicide along with the reflection Au and Si for the sample B, (e) bright field high resolution XTEM image of one of the nanostructure (sample B) showing inter-diffusion of Au into Si. (f) Depicts a SAD pattern taken at RT for the sample that was annealed in-situ TEM at 400°C(sample G) and then cooled to RT. This shows no gold silicide formation at 400°C.

**Figure 2:** (a) Length distribution(fitted with Gaussian distribution) along <1 1 0> direction, of the sample shown in Fig 1 (c) which shows an average length of 48.1± 1.1 nm and (b)corresponding aspect ratio distribution and average aspect ratio is 0.99 ± 0.01 ( here the histograms are fitted with Gaussian distribution to estimate the mean aspect ratio) (c) length distribution(fitted with Gaussian distribution) of the sample shown in Fig 3(a) (for several frames) which shows the average length is 116 ± 1.8 nm and (d) corresponding aspect ratio distribution and average aspect ratio is 1.12 ± 0.03 ( here the histograms are fitted with log -normal distribution to estimate the mean aspect ratio).

**Figure 3:** (a) SEM image taken at 0º tilt (c) 54º tilt; (b) depicts a XTEM image of epitaxially grown Ge nano structures and (d) for Au-Ge bilobed structures (sample C).

**Figure 4:** (a) Bright field TEM image of one of the bi-lobed Au-Ge nanostructures and (b) corresponding HRTEM images showing the Au-Ge alloy formation at the Au-Ge junction. (c) depicts STEM micrograph (sample C) and (d) STEM-EDS elemental mapping (sample C).

**Figure 5:** SEM image taken at RT after 5.0 nm Ge deposited on Au patterned substrate at (a) 400°C (b) 500°C and (c) 600°C. This shows that no bi-lobed structures are formed at 400° C. Corresponding RBS spectra in (d) show diffusion status of Au and Ge in Silicon at 400°C (sample F1), 500°C (sample F2) and 600°C (sample F3).

**Figure 6:** SEM images showing absence of Au-Ge bilobed structures for (a) sample D and (b) for sample E.

**Table 1:** Summary of the samples is mentioned in the tables.

**Figure 1:** Rath A, *et al.,*

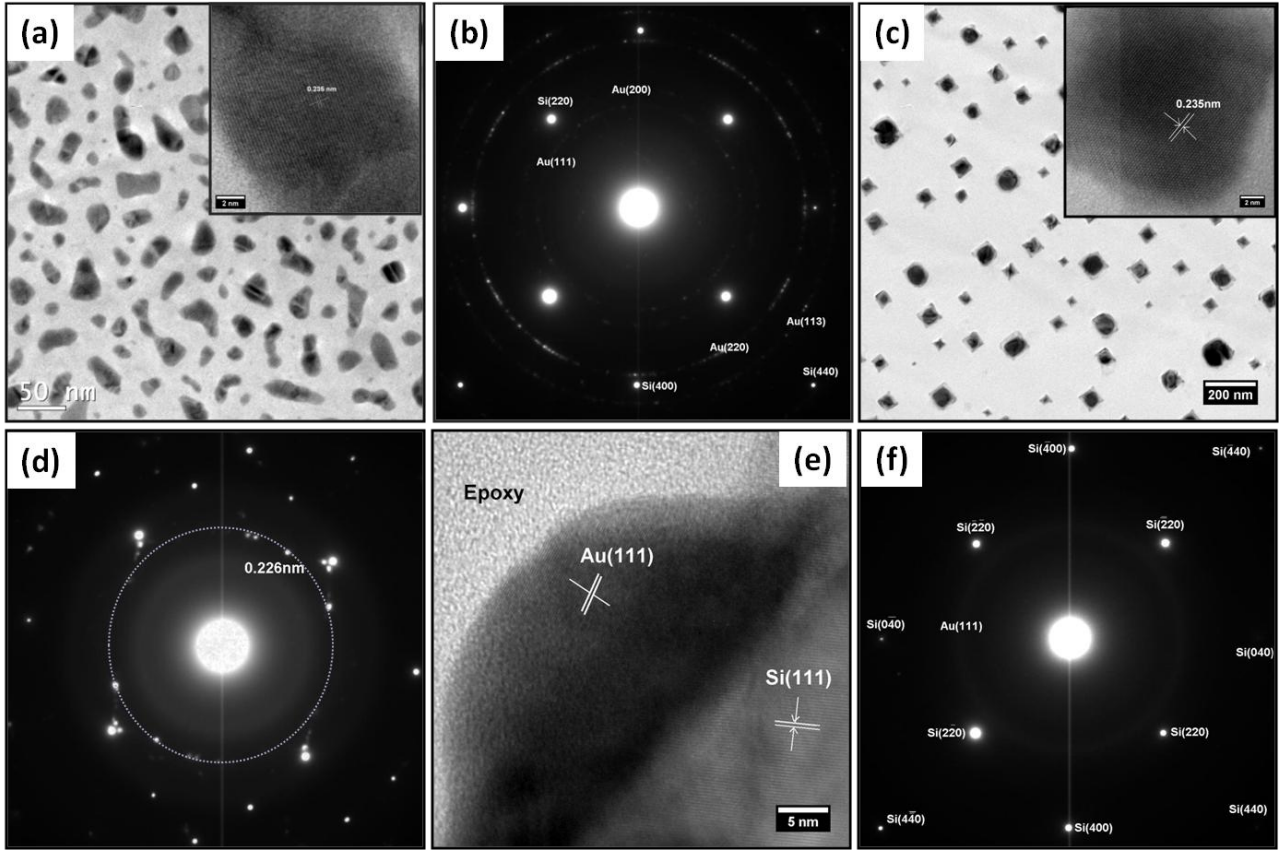

**Figure 2:** Rath A, *et al.,*

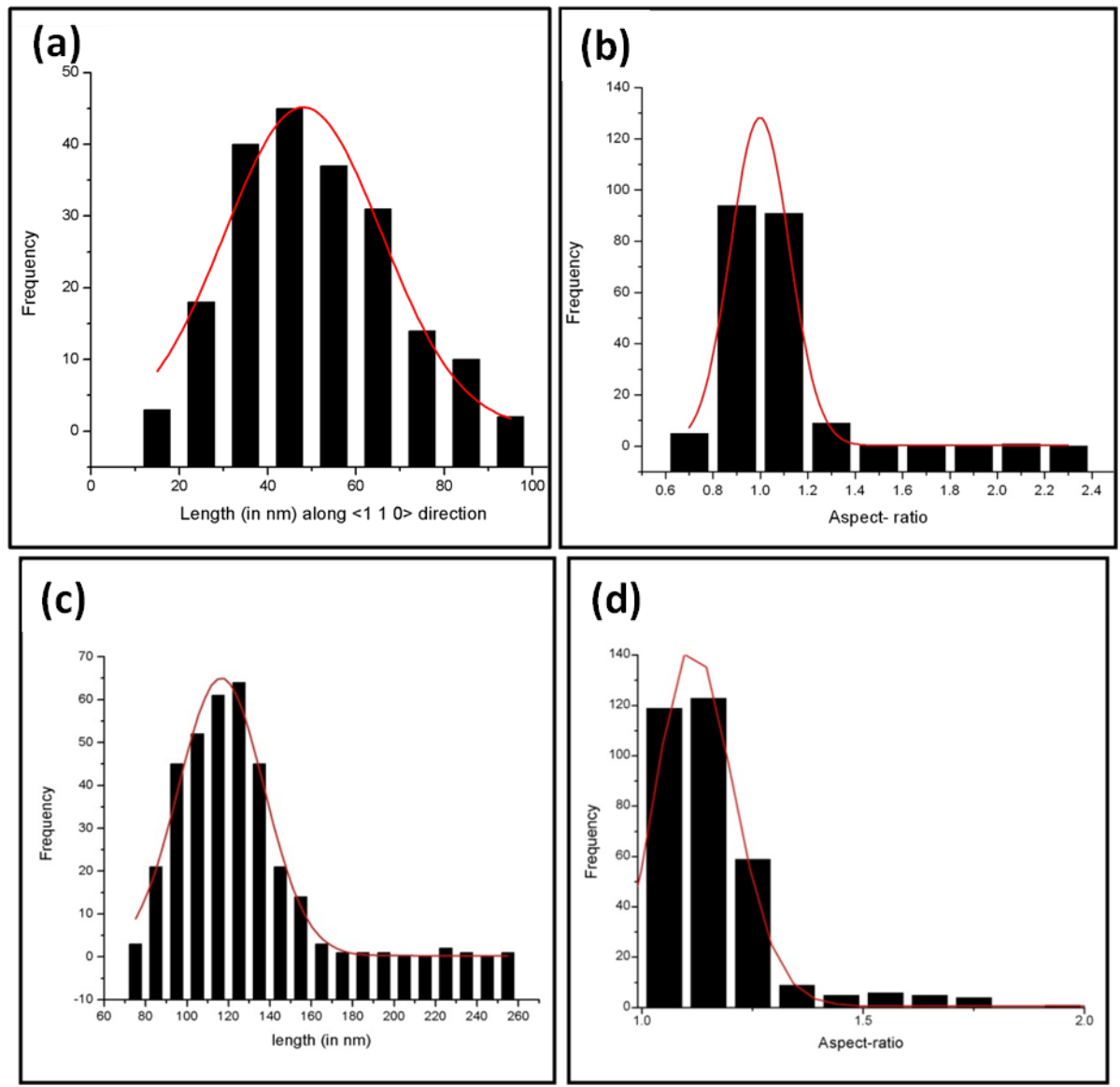

**Figure 3:** Rath A, *et al.,*

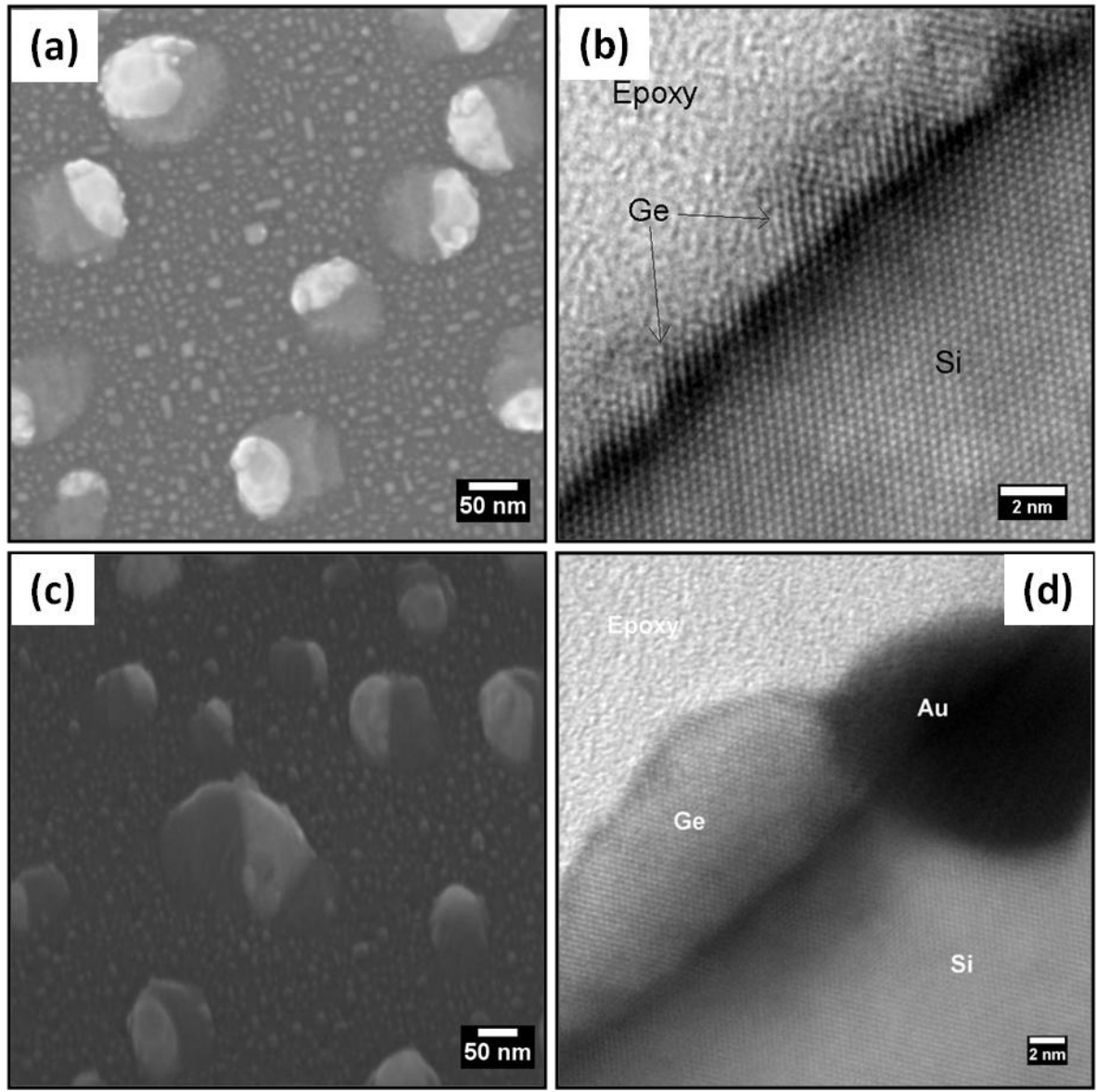

**Figure 4:** Rath A, *et al.,*

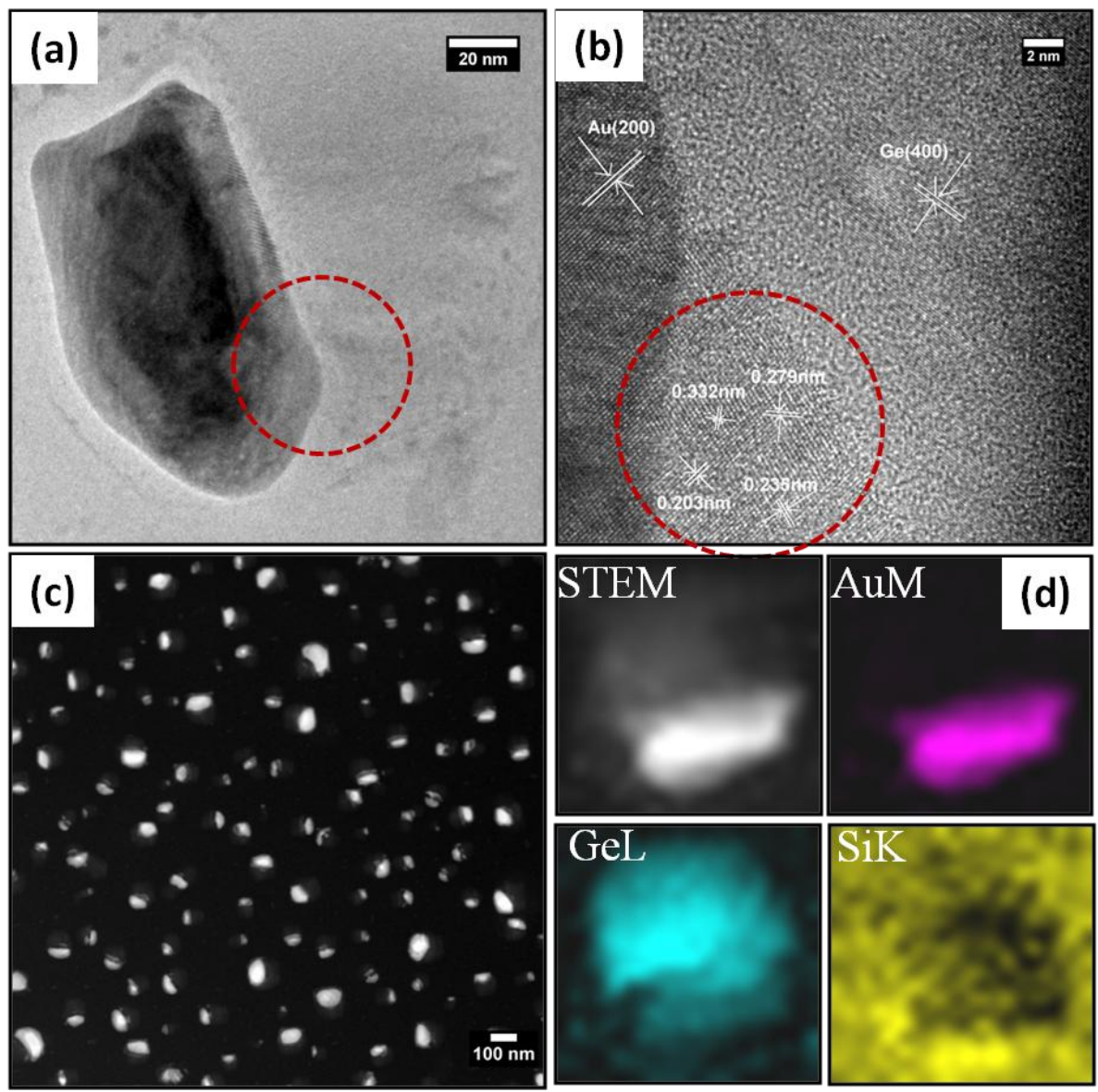

**Figure 5:** Rath A, *et al.,*

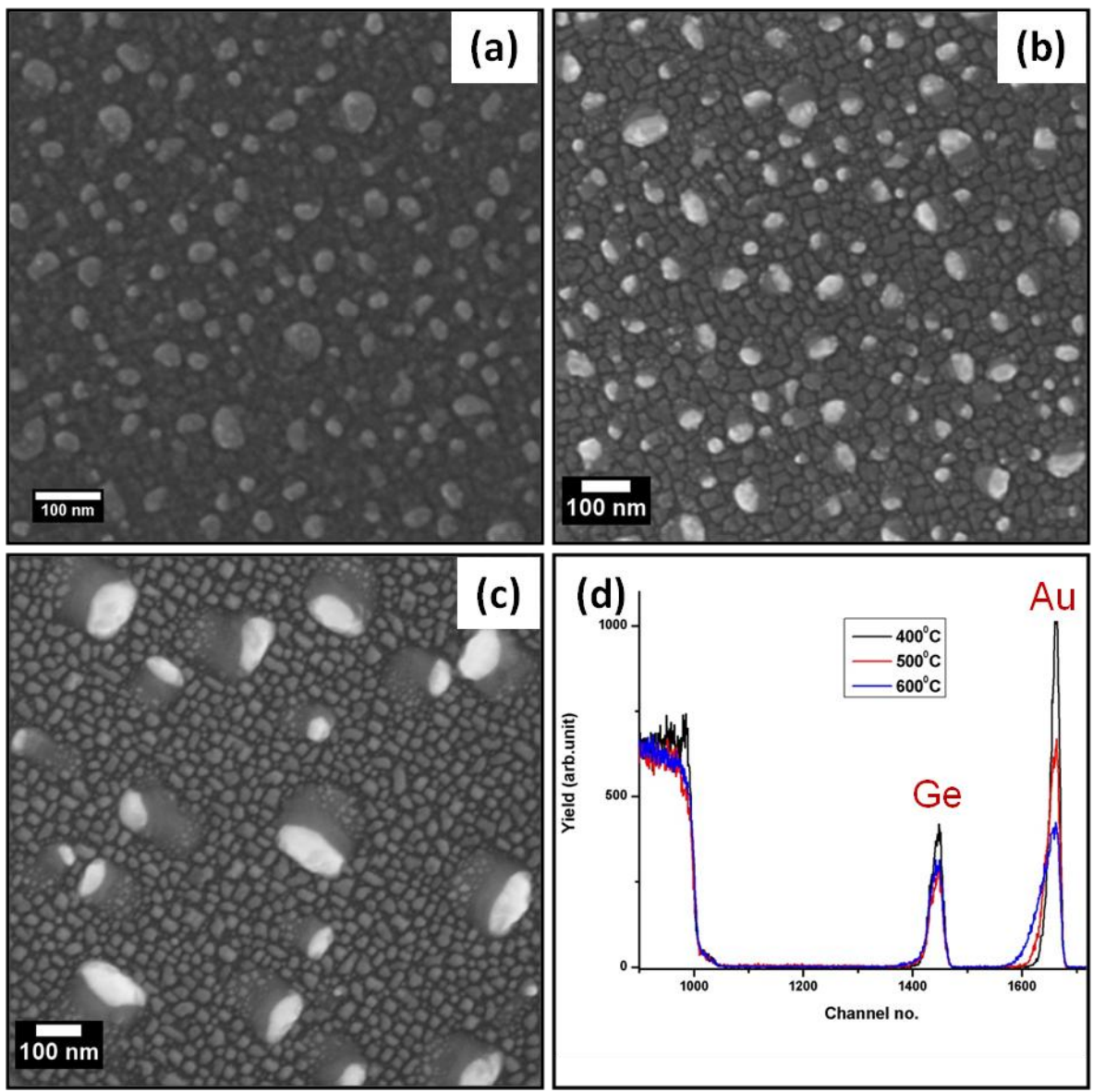

**Figure 6:** Rath A, *et al.,*

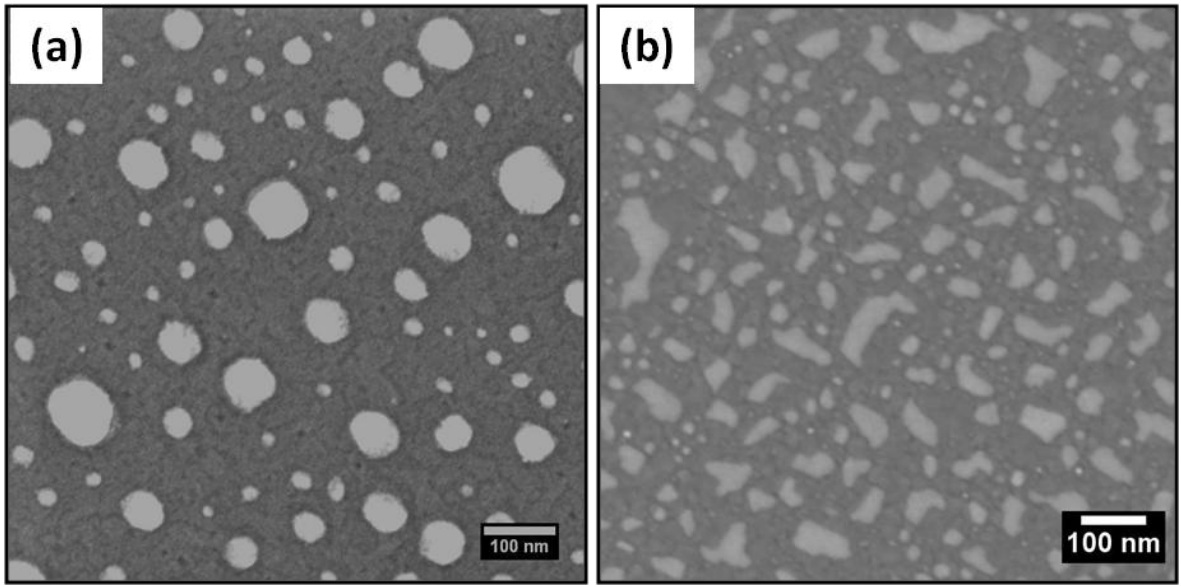

**Table 1:** Rath A, *et al.,*

| Name of the sample | Details of the Experimental conditions |
|---|---|
| Sample A | 2.0 nm Au/Si(100) as-deposited |
| Sample B | 2.0 nm Au/Si(100) annealed at 500º C |
| Sample C | 2.0 nm Ge /Sample B |
| Sample D | 2.0 nm Ge /Sample B @RT(Room Temperature) |
| Sample E | 2.0 nm Ge /2.0 nm Au/Si(100) annealed at 500º C |
| Sample F1 | 5.0 nm Ge /2.0 nm Au/Si(100) annealed at 400º C |
| Sample F2 | 5.0 nm Ge /2.0 nm Au/Si(100) annealed at 500º C |
| Sample F3 | 5.0 nm Ge /2.0 nm Au/Si(100) annealed at 600º C |
| Sample G | Sample A annealed at 400º C inside HV chamber |